\documentclass[useAMS,usenatbib]{mn2e}

\usepackage{graphicx}\usepackage{epsfig}\usepackage{epsf}
\usepackage{amsmath}\usepackage{amssymb}\usepackage{stfloats}
\title[Eta Car's fast ejecta]{Exceptionally fast ejecta seen in light
  echoes of Eta Carinae's Great Eruption}

\author[Smith et al.]{Nathan Smith$^{1}$\thanks{E-mail:
    nathans@as.arizona.edu}, Armin Rest$^{2,3}$, Jennifer
  E.\ Andrews$^1$, Tom Matheson$^4$, \newauthor Federica B.\
  Bianco$^{5,6}$, Jose L.\ Prieto$^{7,8}$, David J.\ James$^9$,
  R.\ Chris Smith$^{10}$, \newauthor Giovanni Maria
  Strampelli$^{2,11}$, and A.\ Zenteno$^{10}$ \\
  $^{1}$Steward Observatory, University of Arizona, 933 N. Cherry
  Ave., Tucson, AZ 85721, USA \\ $^2$Space Telescope Science
  Institute, 3700 San Martin Drive, Baltimore, MD 21218, USA \\
  $^3$Department of Physics and Astronomy, The Johns Hopkins
  University, 3400 North Charles Street, Baltimore, MD 21218, USA \\
  $^4$National Optical Astronomy Observatory, Tucson, AZ 85719, USA \\
  $^5$CCPP, New York University, 4 Washington Place, New York, NY
  10003, USA \\ $^6$Center for Urban Science and Progress, New York
  University, 1 MetroTech Center, Brooklyn, NY 11201, USA \\
  $^7$N\'ucleo de Astronomia de la Facultad de Ingenieria, Universidad
  Diego Portales, Av. Ejercito 441, Santiago, Chile \\ $^8$Millennium
  Institute of Astrophysics, Santiago, Chile \\ $^9$Event Horizon
  Telescope, Smithsonian Astrophysical Observatory MS 42,
  Harvard-Smithsonian Center for Astrophysics, 60 Garden \\ Street,
  Cambridge, MA 02138, USA \\ $^{10}$Cerro Tololo Inter-American
  Observatory, National Optical Astronomy Observatory, Colina El Pino
  S/N, La Serena, Chile \\ $^{11}$Universidad de La Laguna, Tenerife,
  Spain}

\begin{document}

\pagerange{\pageref{firstpage}--\pageref{lastpage}} \pubyear{2012}
\maketitle
\label{firstpage}

\begin{abstract}

  In our ongoing study of $\eta$~Carinae's light echoes, there is a
  relatively bright echo that has been fading slowly, reflecting the
  1845-1858 plateau phase of the eruption. A separate paper discusses
  its detailed evolution, but here we highlight one important result:
  the H$\alpha$ line in this echo shows extremely broad emission wings
  that reach $-$10,000~km~s$^{-1}$ to the blue and
  $+$20,000~km~s$^{-1}$ to the red.  The line profile shape is
  inconsistent with electron scattering wings, so the broad wings
  indicate high-velocity outflowing material. To our knowledge, these
  are the fastest outflow speeds ever seen in a non-terminal massive
  star eruption. The broad wings are absent in early phases of the
  eruption in the 1840s, but strengthen in the 1850s.  These speeds
  are two orders of magnitude faster than the escape speed from a warm
  supergiant, and 5--10 times faster than winds from O-type or
  Wolf-Rayet stars.  Instead, they are reminiscent of fast supernova
  ejecta or outflows from accreting compact objects, profoundly
  impacting our understanding of $\eta$~Car and related transients.
  This echo views $\eta$~Car from latitudes near the equator, so the
  high speed does not trace a collimated polar jet aligned with the
  Homunculus.  Combined with fast material in the Outer Ejecta, it
  indicates a wide-angle explosive outflow.  The fast material may
  constitute a small fraction of the total outflowing mass, most of
  which expands at $\sim$600 km s$^{-1}$.  This is reminiscent of fast
  material revealed by broad absorption during the presupernova
  eruptions of SN~2009ip.

\end{abstract}


\section{INTRODUCTION}

The massive evolved star $\eta$ Carinae serves as a tremendous
reservoir of information about episodic mass loss in the late-stage
evolution of massive stars.  It is uniquely valuable because it is
nearby, because it underwent a spectacular ``Great Eruption'' event
observed in the mid-19th century, and because we can now observe the
spatially resolved shrapnel of that event with modern tools like the
{\it Hubble Space Telescope} ({\it HST}).  Added to this list is the
recent discovery of light echoes from the 19th century eruption
\citep{rest12}, which now allow us to obtain spectra of light from an
event that was seen directly by Earth-based observers before the
invention of the astronomical spectrograph.  This is similar to
studies of light echoes from historical supernovae (SNe) and SN
remnants in the Milky Way and Large Magellanic Cloud
\citep{rest05a,rest05b,rest08}.

Spectroscopy of these light echoes provides informative comparisons
between $\eta$ Car and extragalactic eruptions.  Based mostly on its
historical light curve \citep{sf11}, $\eta$ Car has been a prototype
for understanding luminous blue variable (LBV) giant eruptions and SN
impostors \citep{smith+11,vdm12}.  Eruptions akin to $\eta$ Car have
been discussed in the context of brief precursor episodes of extreme
mass loss that create circumstellar material (CSM) of super-luminous
Type~IIn supernovae \citep{smith+07,sm07}.  In addition to extreme
$\eta$ Car-like mass loss, several lines of evidence connect LBVs and
SNe with dense CSM (see \citealt{smith14} for a review; also e.g.,
\citealt{gl09,groh13,groh14,justham14,kv06,mauerhan13,so06,smith07,smith+08,smith+11,trundle08}).
LBVs have the highest known mass-loss rates of any stars before death,
where LBV giant eruptions can lose as much as several $M_{\odot}$ in a
few years (see \citealt{smith14}).

The physical trigger and mechanism of these LBV-like giant eruptions
are still highly uncertain.  Eruptive mass loss is usually discussed in
the context of super-Eddington winds
\citep{davidson87,og97,owocki04,os16,q16,so06,vanmarle08}.  This
framework addresses how mass can be lost at such a high rate, but it
does not account for where the extra energy comes from. There are also
(sometimes overlapping) scenarios that have been discussed, involving
binary mergers, stellar collisions, violent common envelope events,
accretion events onto a companion (perhaps including compact object
companions, although this has not been discussed much for $\eta$ Car),
violent pulsations, extreme magnetic activity, pulsational pair
instability eruptions, unsteady or explosive nuclear burning, and wave
driving associated with late nuclear burning phases approaching core
collapse
\citep{jsg89,fuller17,gl12,hs09,ks09,pz16,piro11,qs12,sq14,smith11,sa14,smith+11,smith+16,soker01,soker04,woosley17}.
In any case, a tremendous amount of mass (several $M_{\odot}$) leaves
the star in a brief window of time (a few years), and observational
constraints on the outflow properties provide a key way to guide
theoretical interpretation.

In general, quasi-steady radiation-driven winds are expected to leave
a star with a speed that is within a factor of order unity compared to
the escape speed from the star's surface. That is why red supergiant
winds are slow (10s of km s$^{-1}$), blue supergiant winds are a few
hundred km s$^{-1}$, O-type stars have winds around 1000 km s$^{-1}$,
and more compact H-poor Wolf-Rayet star winds are 2000-3000 km
s$^{-1}$ (see \citealt{smith14} for a review).  For example,
line-driven winds of hot O-type stars have a ratio of their terminal
wind to the star's escape speed of $v_{\infty}/v_{\rm esc} \approx
2.6$, and cooler stars below about 21,000 K have $v_{\infty}/v_{\rm
  esc} \approx 1.3$ \citep{lamers95,vink99}.  Line driven wind theory
and observations indicate that $v_{\infty}/v_{\rm esc} \approx 2.6$ is
expected to become lower as the star's temperature drops
\citep{cak,abbott82,pauldrach86,pp90,lamers95,vink99}.  For a strongly
super-Eddington wind in an LBV, where $\Gamma$ substantially exceeds
1, the effective gravity is low, the stellar envelope may inflate, the
wind may show a complicated pattern of outflow and infall, and
material may ultimately leak out slowly.  The atmosphere may be porous
\citep{owocki04}, perhaps leading to a range of outflow speeds, but we
don't expect a steady wind-driven outflow with a high mass-loss rate
to be many times faster than a star's escape speed
\citep{owocki04,vanmarle08,owocki17}.  Numerical simulations of
super-Eddington continuum-driven winds predict terminal wind speeds
below the star's surface escape speed \citep{vanmarle08,vanmarle09}.

Observationally, a wide range of outflow speeds are seen in the $\eta$
Car system.  The bulk outflow of the present-day wind is around
400-500 km s$^{-1}$ \citep{hillier01}, although with some faster
speeds up to around 1000 km s$^{-1}$ in the polar wind
\citep{smith+03}.  The bipolar Homunculus nebula, which contains most
of the mass ejected in the 19th century eruption
\citep{morse01,smith17} has a range of speeds that vary with latitude
from 650 km s$^{-1}$ at the poles to about 50 km s$^{-1}$ in the
pinched waist at the equator \citep{smith06}.  However, there is also
faster material in the system. Hard X-ray emission from the
colliding-wind binary suggests that a companion star has a very fast
wind of 2000-3000 km s$^{-1}$
\citep{corcoran01,pc02,parkin11,russell16}.  In spectra of the central
star system, speeds as fast as $-$2000 km s$^{-1}$ are only seen in
absorption at certain phases, attributed to the companion's wind
shocking the primary star's wind along our line of sight
\citep{groh10}.

So far, the fastest material associated with $\eta$ Car has been seen
in the Outer Ejecta, where filaments have speeds based on Doppler
shifts and projection angles as high as 5000 km s$^{-1}$
\citep{smith08}.  (Most of the Outer Ejecta seen in images are slower,
moving at a few hundred km s$^{-1}$; \citealt{kiminki16,weis12}.) This
fast material, combined with the high ratio of kinetic energy to total
radiated energy in the eruption \citep{smith03}, has led to
speculation that the Great Eruption may have been partly caused by a
hydrodynamic explosion \citep{smith06,smith08,smith13}.  Spectra of
$\eta$ Car's light echoes \citep{rest12,prieto14} also seemed
inconsistent with traditional expectations for a simple wind
pseudo-photosphere \citep{davidson87}, although \citet{os16} showed
that proper treatment of opacity and radiative equilibrium in such a
wind may lead to cool temperatures around 5000~K.  Well-developed
models for sub-energetic and non-terminal explosive events do not yet
exist, but an explosive ejection of material and a surviving star
might arise if energy is deposited in the star's envelope that is less
than the total binding energy of the core, but enough to unbind the
outer layers.  \citet{dessart10} explored how stellar envelopes might
respond to such energy deposition, and found some cases with partial
envelope ejection and model light curves reminiscent of SN
impostors. \citet{rm17} argued that any deep energy deposition at a
rate that substantially exceeds the steady luminosity of the star is
likely to steepen to a shock.  For the specific case of $\eta$~Car's
eruption, \citet{smith13} argued that an explosive ejection of fast
material interacting with a previous slow wind could account for the
historical light curve and several properties of the Homunculus, where
CSM interaction leads to efficient radiative cooling as in SNe~IIn.

In this paper, we present evidence based on Doppler shifts in light
echo spectra from the Great Eruption, which show that there was in
fact an explosive ejection of very fast material relatively late in
the eruption.  The observed speeds in excess of 10,000 km s$^{-1}$
suggest that a small fraction of the mass was accelerated to very high
speeds by a blast wave, confirming similar conclusions based on fast
nebular ejecta observed around the star at the present epoch
\citep{smith08,smith13}.

%%%%%%%%%%%%%%%%%%%%%%%% TABLE %%%%%%%%%%%%%%%%%%%%%%%%%%%%%%%%%%%%%%%
%\begin{center}
\begin{table}\begin{center}\begin{minipage}{3.1in}
      \caption{Optical Spectroscopy of Light Echo EC2}
\scriptsize
\begin{tabular}{@{}lcccccc}\hline\hline
UT Date     &Tel./Intr.     &grating &$\Delta\lambda$ (\AA) &slit &PA  \\ \hline
%
%2011 Dec 23  &Baade/IMACS f2 &     &1$\farcs$? &...$^{\circ}$  \\
%2012 Mar 18  &Baade/IMACS f2 &     &1$\farcs$? &...$^{\circ}$  \\
%2012 Jun 26  &Baade/IMACS f2 &     &1$\farcs$? &...$^{\circ}$  \\
%2012 Oct 13  &Baade/IMACS f4 &300  &1$\farcs$2 &340$^{\circ}$  \\
%2013 Jan 07  &Clay/MAGE      &ech. &1$\farcs$0 &...$^{\circ}$  \\
%2013 Apr 06  &Baade/IMACS f4 &300  &0$\farcs$7 &270$^{\circ}$  \\
%2014 Jan 07  &Clay/MAGE      &ech. &1$\farcs$0 &...$^{\circ}$  \\
%2014 Feb 05  &Baade/IMACS f4 &1200 &0$\farcs$9 &293$^{\circ}$  \\
%2014 Feb 06  &Baade/IMACS f4 &300  &0$\farcs$9 &293$^{\circ}$  \\
%2014 May 19  &Baade/IMACS f4 &1200 &0$\farcs$7 &293$^{\circ}$  \\
2014 Nov 03  &Gemini/GMOS    &R400 &5000-9200 &1$\farcs$0 &293$^{\circ}$  \\
2015 Jan 20  &Baade/IMACS f4 &1200 &5500-7200 &0$\farcs$7 &293$^{\circ}$  \\
2015 Jan 20  &Baade/IMACS f4 &300  &4000-9000 &0$\farcs$7 &293$^{\circ}$  \\
%2016 Mar 04  &Baade/IMACS f4 &1200 &0$\farcs$7 &293$^{\circ}$  \\
%2015 Mar 04  &Baade/IMACS f4 &1200 &0$\farcs$7 &293$^{\circ}$  \\
%2016 Mar 05  &Baade/IMACS f2 &     &0$\farcs$7 &293$^{\circ}$  \\
\hline
\end{tabular}\label{tab:spec}\end{minipage}
\end{center}
%$^a$...
\end{table}%\end{center}

\section{Optical Spectroscopy}

Following the discovery of light echoes from $\eta$ Carinae
\citep{rest12}, we have continued to follow the slow spectral
evolution of several echoes.  So far in previous papers, we have
discussed the initial spectra and spectral evolution of a close group
of echoes (called ``EC1'') thought to arise from pre-1845 peaks in the
light curve \citep{rest12,prieto14}, but we have monitored a number of
other echo systems as well.

One echo, which we refer to henceforth as ``EC2'', captured our
attention because it was relatively bright in our first-epoch image in
2003, and has faded very slowly over several years since then.  EC2
arises from the reflection off of a cometary shaped dust cloud.  We
infer from its slow rate of fading that EC2 reflects light from the
main plateau of the Great Eruption in 1845-1858; this is explained in
more detail in our companion paper (\citealt{smith18} S18 hereafter),
where we analyze the imaging photometry of the echo.  Based on the
probable time delay of about 160 yr, spectra that we have been
obtaining in the last few years trace light emitted by $\eta$~Car in
its mid-1850s plateau.  From the known geometry of light echo
paraboloids \citep{couderc39} and with a known time delay, one can
deduce the viewing angle for any echo.  EC2's position on the sky is
near the EC1 echoes discussed previously by \citet{rest12} and
\citet{prieto14}.  Like those echoes, EC2 views $\eta$~Car from a
vantage point that is near the equatorial plane of the Homunculus
(within $\sim$10$^{\circ}$), with the uncertainty dominated by the
value of the adopted time delay of the echo.  The fact that this
viewing angle is near the equator and not the polar axis is an
important detail, to which we return later.

The full photometric and spectroscopic evolution of EC2 will be
described in detail in a more lengthy paper referenced above (S18).
Here we focus on just one aspect of these data that stands out as an
important and independent result.  Briefly, EC2 spectra show faint but
extremely broad line wings that indicate faster ejecta than has ever
been seen before in $\eta$~Car or in any other massive star eruption.
The result described below was quite astonishing to us, and profoundly
impacts our understanding of the nature of $\eta$ Car's eruption.
Implications are discussed in the next section.

We obtained low- and moderate-resolution spectra of the EC2 echo on a
number of dates from 2011 to the present, but we only mention a few of
those in this paper.  We obtained one spectrum on 2014 Nov 13 using
the Gemini Multi-Object Spectrograph (GMOS) \citep{hook02} at Gemini
South on Cerro Pach\'{o}n.  Nod-and-shuffle techniques \citep{gb01}
were used with GMOS to improve sky subtraction.  Standard CCD
processing and spectrum extraction were accomplished with
IRAF\footnote{IRAF is distributed by the National Optical Astronomy
  Observatory, which is operated by the Association of Universities
  for Research in Astronomy, Inc., under cooperative agreement with
  the National Science Foundation.}.  The spectrum covers the range
$4540-9250$~\AA \, with a resolution of $\sim9$~\AA.  We used an
optimized version\footnote{\tt https://github.com/cmccully/lacosmicx}
of the LA Cosmic algorithm \citep{vandokkum01} to eliminate cosmic
rays.  We optimally extracted the spectrum using the algorithm of
\citet{horne86}.  Low-order polynomial fits to calibration-lamp
spectra were used to establish the wavelength scale.  Small
adjustments derived from night-sky lines in the object frames were
applied.  We employed our own IDL routines to flux calibrate the data
using the well-exposed continua of spectrophotometric standards
\citep{wade88,matheson00}.

An extensive series of spectra was obtained using the Inamori-Magellan
Areal Camera and Spectrograph (IMACS; \citealt{dressler11}) mounted on
the 6.5m~Baade telescope of the Magellan Observatory. Most of these
spectra are presented in our more lengthy paper, but we selected a few
epochs here to demonstrate the clear detection of the broad line
wings.  The chosen slit width for these observations was 0$\farcs$7.
With the f/4 camera, we used the 300 lines per mm (lpm) grating to
sample a wide wavelength range at moderate $R \simeq 500$ resolution,
and the 1200 lpm grating to sample a smaller wavelength range with
higher resolution of $R \simeq 6000$.  The 1200 lpm grating was
centered on H$\alpha$.  The 2-D spectra were reduced and extracted
using routines in the IMACS package in IRAF.  With the f/4 camera of
IMACS, the spectrum is dispersed across 4 CCD chips, so small gaps in
the spectrum appear at different wavelengths depending on the grating
and tilt used.

We correct all our spectra for a reddening of $E(B-V)$=1.0 mag.  This
includes the minimum line-of-sight reddening in the interstellar
medium (ISM) toward the Carina Nebula of $E(B-V)$=0.47 mag
\citep{walborn95,smith02}, plus an additional $\sim$0.5 mag to account
for extra extinction from dust clouds within the southern part of the
Carina Nebula.  This choice is explained in more detail in S18, but
briefly, this reddening allows the continuum shape of various echoes
to match spectral diagnostics of the temperature \citep{rest12}.  When
we deredden our spectra, the continuum shape is consistent with a
temperature of about 6000~K.  A different choice of $E(B-V)$ would
alter the inferred continuum temperature, but would not significantly
alter our conclusions about the fast ejecta, which are based on line
profile shape.

\begin{figure}
\includegraphics[width=3.4in]{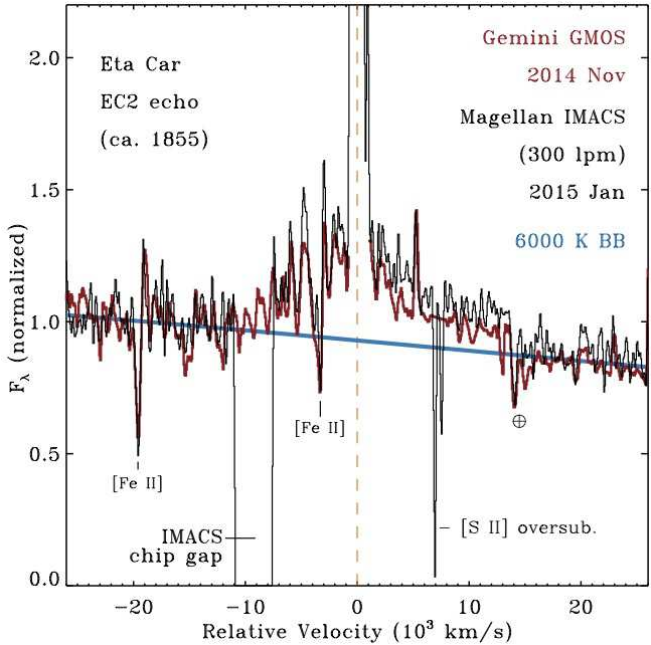}
\caption{Low-resolution spectra of light echo EC2 corresponding to
  late times in the plateau phase of the eruption, showing very broad
  H$\alpha$ line wings.  The relative Doppler shift is plotted in
  units of 10$^3$ km s$^{-1}$, and the broad wings of H$\alpha$ appear
  to extend to at least $\pm$10,000 km s$^{-1}$.  The blue curve shows
  a 6000 K blackbody matched to the continuum.}
\label{fig:fast}
\end{figure}

\begin{figure}
\includegraphics[width=3.4in]{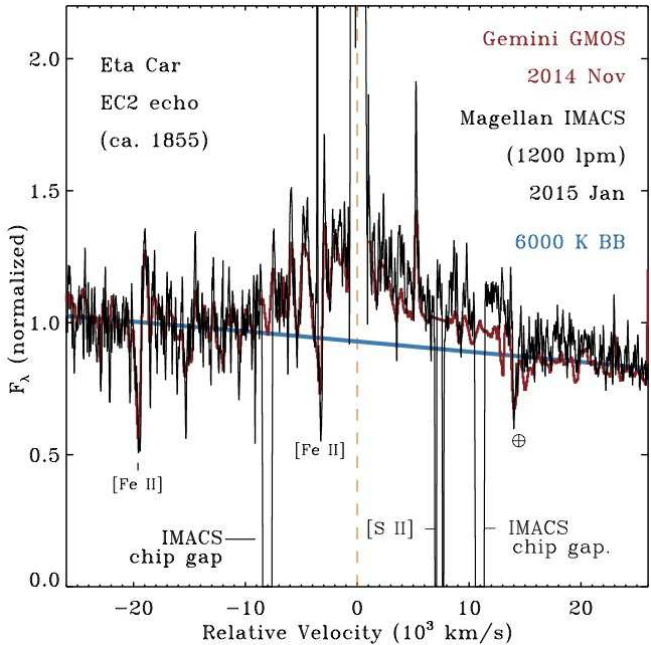}
\caption{Same as Figure~\ref{fig:fast}, but replacing the IMACS
  low-resolution spectrum (300 lpm grating) with a higher-resolution
  spectrum (1200 lpm grating) obtained on the same observing run.  The
  same broad component appears with roughly the same strength even
  though a different grating is used.  The same GMOS spectrum as in
  Figure~\ref{fig:fast} is shown again (in red) for comparion.}
\label{fig:fast1200}
\end{figure}

\begin{figure}
\includegraphics[width=3.4in]{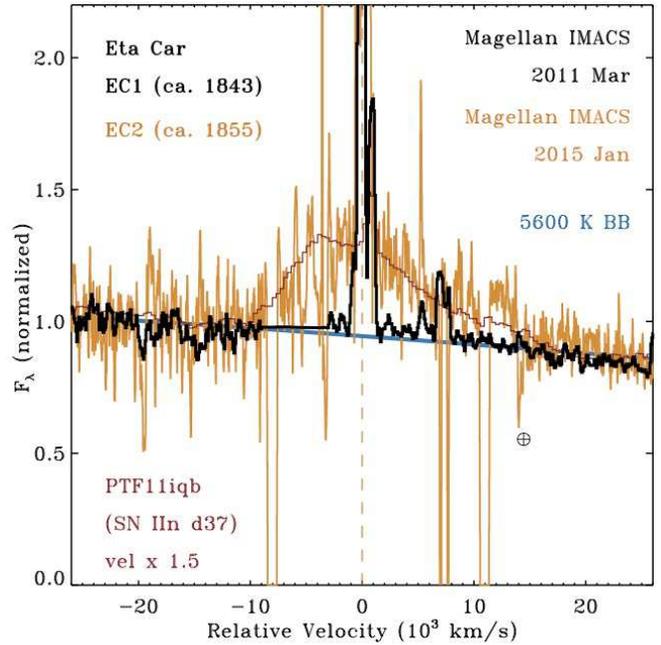}
\caption{Similar to Figures 1 and 2, but here we plot the
  higher-resolution EC2 spectrum (1200 lpm grating) from
  Figure~\ref{fig:fast1200} in orange.  In black, we show the spectrum
  of a different echo (EC1), discussed previously
  \citep{rest12,prieto14}, which reflects light from an early peak in
  the Great Eruption (1843 or 1838), also obtained with IMACS.  This
  echo that traces an earlier phase of the eruption {\it does not}
  show the broad wings, so the fast material seems to have appeared at
  late phases in the eruption. Merely as an illustrative comparison,
  we also show the broad component in the Type IIn/II-L core-collapse
  supernova PTF11iqb from \citet{smith15}, with velocities multiplied
  by 1.5.}
\label{fig:fast1843}
\end{figure}

\section{Results and Discussion}

\subsection{Broad H$\alpha$ Wings Trace Fast Material}

Examining our low-resolution spectra of EC2, we noticed that the
continuum shape in raw spectra appeared to have an unusual kink around
H$\alpha$ at some epochs.  After flux calibration and correcting for a
modest amount of reddening of $E(B-V)$=1.0 mag, this turned out to be
low-level excess emission above the smooth continuum level.  Our IMACS
spectra all have chip gaps near H$\alpha$, so at first we were
suspicious of a relative flux calibration offset on adjacent
chips. Then in 2014 we obtained a low-resolution spectrum of EC2 with
Gemini GMOS, which has no detector chip gap, and we saw the same
excess.  Subsequent low-resolution IMACS spectra show that this excess
appeared to be growing in strength with time, more so on the red wing.

Figure~\ref{fig:fast} shows the low-resolution spectrum of EC2 near
H$\alpha$, taken in 2014 with Gemini/GMOS (red-orange) and in 2015
with Magellan/IMACS (black) using the low-resolution 300 lpm grating.
These epochs of the echo correspond to a date during the eruption in
the mid-1850s.  Broad emission wings of H$\alpha$ are seen in both
spectra, and although the signal to noise in the continuum is low, the
two spectra agree quite well -- except that the broad emission is
perhaps a bit stronger in 2015.  The emission wings are interrupted by
Fe~{\sc ii} P Cygni profiles intrinsic to $\eta$ Car, as well as
oversubtracted nebular [S~{\sc ii}] emission, telluric absorption (the
B band), and the IMACS chip gap.

Reducing and extracting the spectra of faint light echoes can be a bit
tricky, especially in cases like $\eta$ Car where we have extremely
faint reflected light in the echo that is embedded in a bright H~{\sc
  ii} region.  To double check that the faint broad emission wings are
not due to instrumental scattering from the bright nebular H$\alpha$
emission from the Carina Nebula H~{\sc ii} region included in the same
slit (which was, in principle, carefully subtracted with the sky
emission by sampling adjacent regions along the slit), we also
examined our higher resolution spectra of EC2 taken with IMACS using
the 1200 lpm grating.  At first glance, the broad wings were not
apparent in our higher-resolution spectra, but this is because the
wings are so broad that they are dispersed thinly across almost the
entire sampled wavelength range and they look like continuum.  In
fact, the same broad wings are seen clearly in the higher resolution
1200 lpm grating IMACS spectra when they are plotted on the same
intensity scale as the low-resolution spectra.
Figure~\ref{fig:fast1200} is the same as Figure~\ref{fig:fast}, except
that we replaced the 300 lpm IMACS spectrum with the 1200 lpm IMACS
spectrum obtained on the same night.  The two IMACS spectra are
consistent within the limitations of signal to noise.  Since we see
the same broad emission wings in two different instruments on two
different telescopes, and also with two different gratings in the same
instrument, this broad emission must be real and intrinsic to $\eta$
Carinae's light echo.

\begin{figure}
\includegraphics[width=3.1in]{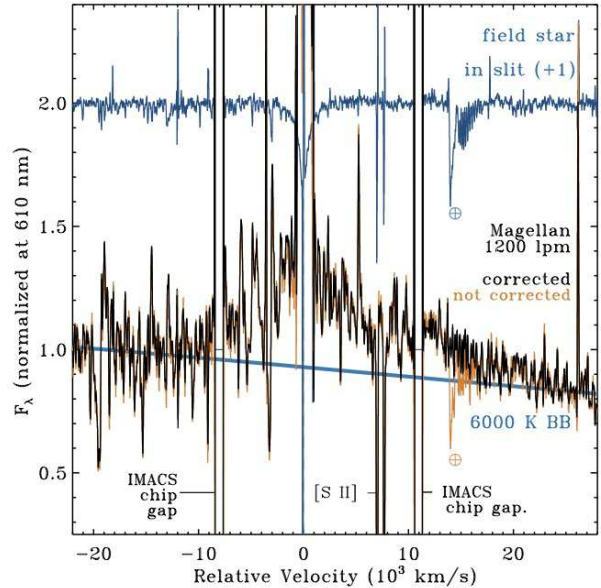}
\caption{The effects of telluric absorption on the broad redshifted
  emission wing of H$\alpha$ in the EC2 spectrum.  The orange tracing
  (with no correction for telluric absorption) is the same IMACS 1200
  lpm spectrum of EC2 as in Figure~\ref{fig:fast1200}.  The black
  tracing shows this same H$\alpha$ line profile after correcting for
  telluric absorption by the B-band (marked by $\earth$). The blue
  tracing is the spectrum of a field star that was included about one
  arcminute away in the same slit, in the same exposures as the
  observations of EC2.  This is not a proper telluric standard, and
  the strong H$\alpha$ absorption corrupts the H$\alpha$ emission of
  EC2 at low velocities, but it is useful to correct for the telluric
  absorption of the B-band on the red wing of the line.  After
  correcting for telluric absorption (black) the red wing drops
  smoothly from $+$12,000 km s$^{-1}$ to beyond $+$20,000 km s$^{-1}$.
  Exactly how much farther the red wings extends beyond $+$20,000 km
  s$^{-1}$ depends on the choice of the admittedly noisy continuum
  level.}
\label{fig:tell}
\end{figure}

These broad emission wings are not due to some other source of broad
emission within the Carina Nebula, because the broad wings are absent
in spectra of other (EC1) light echoes from $\eta$ Car in the same
part of the sky (about 2$\arcmin$ away), which trace earlier peaks
(1843 or before) in the Great Eruption \citep{rest12,prieto14}.  A
spectrum of one of these earlier EC1 echoes (also obtained with the
IMACS spectrograph on Magellan) is shown in Figure~\ref{fig:fast1843}
(thick black line), as compared to the same 1200 lpm spectrum of EC2
(orange).  {\it Clearly the broad component is absent or much weaker
  in the EC1 echo that corresponds to an earlier phase of the
  eruption.}
%See Figure 2 in \cite{prieto14}, which includes spectra of echoes from
%those early peaks taken with Gemini/GMOS and Magellan/IMACS; those
%spectra do not show the broad emission wings we report here.  
This confirms that the broad emission wings appeared relatively late
in the Great Eruption.  Even in our spectra of this new EC2 echo, the
broad wings are somewhat weaker in our earliest spectra in 2011/2012,
but strengthen through 2014 and 2015 until the present.  From the
adopted time delay for the echo of $\sim$160 years, this suggests that
the broad emission wings became prominent in the mid-1850s.  It is
interesting that the emergence of fast material in spectra occurs
shortly after the ejection date of 1847.1 ($\pm$0.8 yr) deduced from
proper motions of the Homunculus \citep{smith17}.  The broad emission
wings are also absent in the extracted spectrum of a field star that
was included in the same IMACS slit aperture about an arcminute away
from EC2 (Figure~\ref{fig:tell}), confirming that they are not
instrumental.

%\footnote{Note that the time it takes for the broad wing emission to
%  appear and fade (a few years) is not necessarily the true duration
%  over which the fast ejecta were present.  This is because any echo
%  signal is smeared in time due to the light travel time across the
%  reflecting surface, which in this case is a cometary cloud with a
%  size of 1--2 ly.  Thus, it is possible that the appearance of the
%  broad wings was a brief event in the mid-1850s, although the true
%  duration is difficult to constrain from data.}

The broad H$\alpha$ emission wings in Figures~\ref{fig:fast} and
\ref{fig:fast1200} are unprecedented for an LBV eruption.  The blue
wing extends to $-$10,000 km s$^{-1}$.  The red wing extends past
+10,000 km s$^{-1}$, to at least +14,000.  It is difficult to
determine the farthest extent of the red wing due to the telluric B
band absorption (marked by $\earth$ in the figures) overlapping that
part of the line profle.  Figure~\ref{fig:tell} presents a correction
for the telluric absorption on the red wing of H$\alpha$.  Although we
did not obtain an appropriate set of telluric standard star
observations to correct the full spectra, one of our slit positions
using the IMACS 1200 lpm grating to observe EC2 also included a bright
field star in the same slit about an arcminute away.  This star
appears to be a late B-type or early A-type star, so the H$\alpha$
absorption intrinsic to this star is fairly strong and corrupts the
telluric correction at low velocities ($\pm$1500 km/s). However, the
continuum in the vicinity of the atmospheric B-band is smooth and
provides a suitable way to correct for the telluric absorption on the
red wing of H$\alpha$.  The extracted spectrum of this field star is
shown in blue in Figure~\ref{fig:tell}.  We divided the observed
spectrum of EC2 by this field star to correct for the telluric B-band
absorption.  Figure~\ref{fig:tell} shows the IMACS 1200 lpm spectrum
before (orange) and after (black) correction of the telluric
absorption.  After correcting for telluric absorption in this way, the
resulting line profile shows that the broad wings are asymmetric,
extending farther on the red side out to at least $+$20,000 km
s$^{-1}$.  Both the red and blue wings are extreme compared to the
relatively narrow core of the H$\alpha$ line, which has a width of
about $\pm$600 km s$^{-1}$.

\begin{figure}
\includegraphics[width=3.1in]{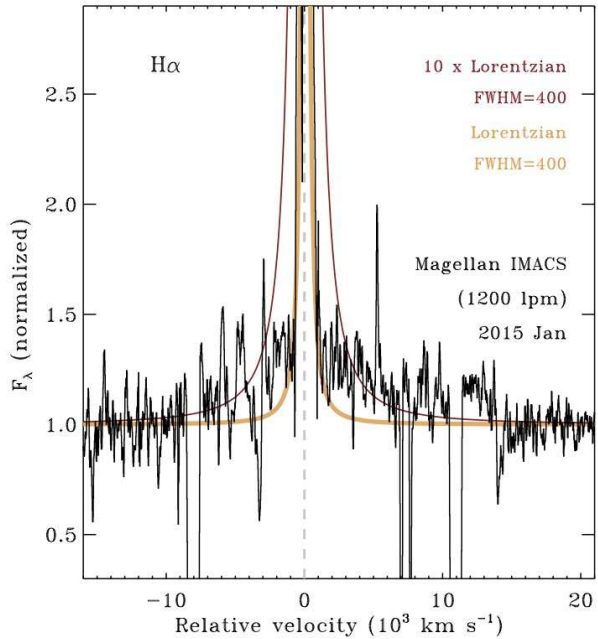}
\caption{The observed broad wings compared to Lorentzian profiles
  expected for electron scattering wings.  The black spectrum is the
  same Magellan/IMACS 1200 lpm spectrum of EC2 in
  Figure~\ref{fig:fast1200}, normalized to the red continuum.  If the
  broad wings were caused by electron scattering, they should follow a
  roughly symmetric Lorentzian profile.  The width of any such profile
  is restricted by the width of the narrow line core at about 1.5
  times the continuum level in this plot.  The thick orange curve
  shows a Lorentzian profile with FWHM = 400 km s$^{-1}$ that roughly
  approximates the width at the base of the central line core, and it
  vastly underpredicts the flux of the broad wings and has the wrong
  shape.  The thinner red curve is the same Lorentzian multiplied by a
  factor of 10 in flux above the continuum; this can account for the
  flux in the very broad wings, but then the Lorentzian profile vastly
  overpredicts the flux at around $\pm$1500 km s$^{-1}$. }
\label{fig:lorentzian}
\end{figure}

Narrow line cores can have broad electron-scattering wings, as is
commonly seen in dense stellar winds and early spectra of SNe~IIn, and
these wings can extend to larger velocities than the true kinematic
speed of ejecta.  Indeed, the present-day wind spectrum of $\eta$ Car
shows an H$\alpha$ profile with electron-scattering wings that extend
beyond $\pm$1000 km s$^{-1}$ \citep{hillier01,smith+03}, even though
the wind speed is only about 400-500 km s$^{-1}$.  However, electron
scattering cannot explain the broad emission wings seen in spectra of
EC2, because they are too broad and have the wrong shape.  With a
characteristic temperature around 6000~K, scattering off thermal
electrons will produce wings with a width of only about $\pm$500-1000
km s$^{-1}$.  Even in the early phases of SNe~IIn when the
temperatures might be 20,000~K (as in $\eta$ Car's present-day wind),
the electron scattering wings only have widths (FWHM) of typically
$\pm$1500 km s$^{-1}$.  Moreover, the broad wings seen in
Figures~\ref{fig:fast} and \ref{fig:fast1200} do not have the
characteristic symmetric Lorentzian profiles expected for electron
scattering.

Figure~\ref{fig:lorentzian} compares the observed H$\alpha$ profile to
Lorentzian profile shapes.  A Lorentzian profile is expected if the
wings are produced primarily by electron scattering, as is seen in
early observations and models of SNe~IIn
\citep{chugai01,smith08,dessart16}. It is clear that a Lorentzian
profile cannot match both the broad wings and the narrow width at the
base of the central narrow emission line component in $\eta$~Car's
echo.  The thick orange Lorentzian, with FWHM = 400 km s$^{-1}$,
roughly matches the width at the base of the narrow component (note
that this is not a good approximation of the shape of the narrow
component, which is somewhat irregular and asymmetric, but it does
agree with the widest point at the base of the narrow component).
However, when this profile is extrapolated to high velocities, it
vastly underpredicts the observed flux in the broad wings.  If we
increase the flux of this Lorentzian (thin red profile) so that it can
produce the observed flux in the broad wings at $\pm$5000 km s$^{-1}$,
we find that it vastly overpredicts the flux at lower speeds around
1000-2000 km s$^{-1}$, so that the base of the narrow component is
much broader than observed.  This Lorentzian profile also vastly
overpredicts the total H$\alpha$ line flux.  We conclude that electron
scattering due to high optical depths in a slower outflow cannot
explain the broad wings we report in light echoes of $\eta$~Car.

Instead, the broad emission must trace fast ejecta, separate from the
slower material emitting the narrower line core.  The observed broad
component has a peak shifted to the blue, with a very long red tail.
This asymmetry might be due to partial P Cygni absorption of the broad
blue emission wing.  Similar broad profiles with blue peaks and
extended red wings are sometimes seen in fast ejecta from SNe; an
example from the SN II-L/IIn PTF11iqb \citep{smith15} is shown in red
in Figure~\ref{fig:fast1843}.  In this comparison (meant to illustrate
a possibly similar line-profile shape), we have multiplied the outflow
velocity of PTF11iqb by a factor of 1.5 for comparison, because the
wings in $\eta$ Car's echo are actually broader than in this
core-collapse SN.

\begin{figure*}
\includegraphics[width=4.7in]{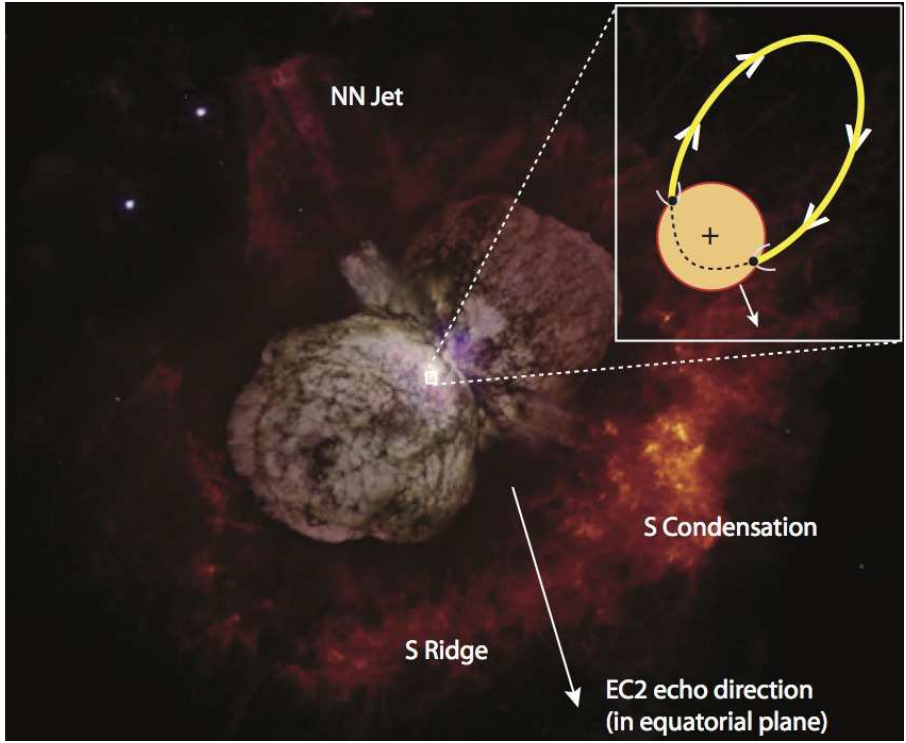}
\caption{A color {\it HST}/WFPC2 image of $\eta$ Car and its ejecta
  (N. Smith/U. Arizona and NASA) showing the direction from which the
  EC2 light echo views $\eta$ Car (this is in or near the equatorial
  plane).  The diagram inset at the upper right shows the orbital
  orientation of the eccentric binary system adapted from
  \citet{madura12}, but with the primary star (orange) bloated to a
  large radius appropriate for its temperature and luminosity in the
  eruption, adopted from \citet{smith11}.  The arrows show the
  direction of motion of the secondary, the two black dots are the
  points of ingress and egress, and the dashed curve is the part of
  the orbit when the secondary is inside the primary star's bloated
  envelope.  The directions of ingress and egress are curiously
  similar to the trajectories of the S~Condensation and NN Jet,
  respectively.}
\label{fig:collision}
\end{figure*}

\subsection{Implications}

The broad wings of H$\alpha$ seen in light echo spectroscopy of $\eta$
Car are remarkable, and are reminiscent of SN ejecta speeds.  To our
knowledge, these are the fastest speeds yet seen in any non-terminal
massive star eruption.  The fastest speeds seen from the central star
(as fast as $-$2000 km s$^{-1}$) are only seen in absorption at
certain phases, and have been attributed to the companion's wind as it
shocks the primary star's wind along our line of sight \citep{groh10}.

The broad emission wings that we report are faint, and only trace a
fraction of the total outflowing matter.  It is not necessarily an
insignificant fraction, though; the integrated flux of the broad
component is about 1/3 of the total H$\alpha$ line flux.  It appears
relatively faint because it is spread across such a wide wavelength
range.  The brighter narrow emission line core of H$\alpha$ shows
widths around 500-600 km s$^{-1}$, more in line with the bulk outflow
velocity in the Homunculus nebula.  Even if it is only tracing a
portion of the mass, an outflow speed of $\sim$10,000 km s$^{-1}$ or
more is surprising, and has important implications for understanding
the basic nature of $\eta$ Car's eruption and its energy budget.  There
are a few key considerations:

1.  The outflow speeds of 10,000 - 20,000 km s$^{-1}$ indicated by the
broad H$\alpha$ wings are two orders of magnitude faster than the
escape speed from a warm supergiant.  For $\eta$ Car in its cool,
bloated eruption state (a radius of a few hundred $R_{\odot}$;
\citealt{smith11}), we would expect a dense, super-Eddington wind to
have a speed on the order of the escape velocity or less
\citep{owocki04,vanmarle08,vanmarle09,smith13}, or about 200 km
s$^{-1}$.  This is indeed the speed observed in light echoes from the
early 1840s peaks in the eruption \citep{rest12,prieto14}, and the
speed observed in spectra of the 1890 event \citep{walborn+liller}.
The broad lines that developed in $\eta$~Car's echo indicate that
something besides a steady wind is at work.  The extremely fast speeds
are also $\sim$10 times faster than the escape speed from an O-type
star or Wolf-Rayet star, which might correspond to $\eta$ Car's
companion star that survives today.  Instead, such high speeds are
reminiscent of outflows from disks around compact objects or fast
shock-accelerated ejecta in SN explosions.  The relatively late
emergence of the fast material may be consistent with the suggestion
\citep{smith13} that a blast wave would accelerate as it encountered a
steep drop in density upon existing the outer boundary of a dense CSM
shell, or it may be indicative of the fastest ejecta getting excited
near a reverse shock during later CSM interaction.

%compact object?

2. The fact that this echo reflects light seen from a vantage point
near the equatorial plane of the Homunculus is critical to
interpreting the broad emission.  Fast outflow velocities could be
interpreted as evidence of a jet that arises when material is accreted
onto a companion star \citep{ks09,soker01,soker04,ts13}.  In such
models, it has been proposed that a main-sequence O-type star accretes
matter at periaston passages and blows bipolar jets that shape the
Homunculus Nebula.  However, it is difficult to see how such a jet
could achieve speeds well in excess of the escape speed from an O-type
dwarf; the observed speeds might be more indicative of
accretion-driven jets from a compact object companion.  More
importantly, however, even this jet scenario would not result in such
fast outflow speeds in the {\it equatorial} direction, because in such
models, a highly collimated polar jet is invoked to shape the bipolar
Homunculus lobes.  That polar axis is perpendicular to the viewing
angle of the echo discussed here. Instead, the broad wings in echoes
seen from low latitudes point to something more like a wide-angle
explosion.

3.  The fast material indicated by the broad wings in the EC2 echo may
be related to fast nebular material seen today in the Outer Ejecta of
$\eta$ Car.  Most of the bright material in the Outer Ejecta is
composed of dense nitrogen-rich condensations moving at speeds of
several hundred km s$^{-1}$
\citep{davidson82,kiminki16,mehner16,sm04,smith08,walborn76,weis01,weis12}.  Some
features seen in images are moving at $\sim$10$^3$ km s$^{-1}$
\citep{sm04,weis01,weis12}.  However, the fastest material can only be
seen in spectra (Doppler shifted out of narrow-band imaging filters);
it appears to be concentrated in polar directions, and is expanding
away from the star at around 5,000 km s$^{-1}$ with a likely origin
during the Great Eruption \citep{smith08}.  It seems probable that the
fastest nebular material in the Outer Ejecta seen today may be a
counterpart to the fast material seen in light echo spectra.  If an
explosion produced ejecta with a range of speeds up to 20,000 km
s$^{-1}$, then the fastest of this material would have already crashed
into the reverse shock, and has therefore given up its kinetic energy
to power the X-ray shell around $\eta$ Car \citep{seward01,sm04}.
Material expanding toward the poles at $\sim$5,000 km s$^{-1}$ or less
is still in free expansion because it has not yet reached the reverse
shock, consistent with its observed location inside the X-ray shell
\citep{smith08}.

Since the EC2 echo discussed here views $\eta$~Car from a latitude
near the equator, it is interesting to speculate about implications
for the equatorial ejecta around the Homunculus and connections to the
central binary system.  The brightest feature in the Outer Ejecta of
$\eta$ Car is the so-called ``S Condensation'' (part of the ``S
Ridge''), located to the S/SW from the star and redshifted
\citep{walborn76,davidson82,kiminki16,sm04}.  Its trajectory of
ejection from the star is not far from the viewing angle of the echo
discussed in this work (see Figure~\ref{fig:collision}).  Is the S
Condensation related to the fast material observed in light echo
spectra?  Today the S~Condensation is much slower (expanding at a few
hundred km s$^{-1}$), but the S Condensation we see today could be the
end product of a small mass of very fast ejecta from the Great
Eruption that swept up and shocked much denser and slower CSM in the
equator.  Proper motions of material in the S Ridge yield ejection
dates several decades before the Great Eruption
\citep{kiminki16,morse01}, consistent with older and slower CSM that
may have been accelerated when hit by the fast ejecta.

Perhaps even more interesting (and more speculative) is a possible
connection to violent binary interaction during the eruption itself.
\citet{smith11} pointed out that with the emitting radius required to
achieve the observed luminosity during the lead-up to the eruption,
the bloated primary star's effective photosphere was actually bigger
than the separation of the two stars at periastron in the current
binary system --- this means that the secondary star plunged into the
extended envelope of the primary (or some extended common envelope
around the system) and came out the other side, doing so multiple
times.  The orientation of the present-day eccentric binary system is
aligned with the equatorial plane of the Homunculus, and has the
secondary star orbiting clockwise on the sky, with periaston in the
direction away from Earth \citep{madura12}.  With this geometry (shown
in Figure~\ref{fig:collision}), the point of ingress when the
secondary star collided with the bloated envelope is on the S/SW side
of the star, such that the direction toward the echo is similar to the
point of ingress.  In other words --- {\it the echo is situated
  favorably to have viewed the secondary star plunging into the
  bloated primary star's envelope}.  (Of course, the central ``primary
star'' here may have been a close binary in the midst of a merger
event surrounded by a common envelope.)  Similarly, the direction of
egress from the bloated envelope seems well aligned with the
trajectory of the NN Jet (Figure~\ref{fig:collision}).  Some
speculation is required here, since to our knowledge there have been
no 3-D hydrodynamic numerical simulations to explore such a scenario
of two massive stars colliding in an eccentric binary with a bloated
primary.  One can imagine a small mass of high velocity ejecta
accelerated by the ensuing splash, which may far exceed the orbital
speed of the secondary, or a small amount of shock acclerated ejecta
escaping through a chimney formed by the wake of the orbiting
companion.  This violent collision may have had something to do with
the fast ejecta revealed in the spectra of light echoes reported here,
and it may have ejected fast material preferentially in the directions
of the S Condensation and NN~Jet.

% We note that the recognition of such extremely fast outflow speeds
% during the Great Eruption was only possible through the use of light
% echoes that allow us to see the spectrum of the eruption itself.
% The fastest material evident from these broad line wings would
% probably be decelerated very quickly as it caught up with and
% shocked slower surrounding material.  While the imprint of that fast
% material may still be evident in the form of kinetic energy that it
% imparted to the surrounding gas, producing the outer ejecta and
% X-ray shell, nebular kinematics alone cannot diagnose the true
% initial ejection speed.

\subsection{SN~2009ip}

There are also interesting connections to extragalactic objects.  Most
LBVs and SN impostors show H$\alpha$ line widths that indicate bulk
outflow speeds from a few hundred up to 1000 km s$^{-1}$ \citep{smith+11},
like the majority of the mass flux in $\eta$ Car.  Similar speeds are
seen in the CSM of Type~IIn supernovae (see \citealt{smith14}).  

One remarkable exception is the case of SN~2009ip, which was a
luminous and eruptive LBV-like star observed in eruption in 2009
\citep{smith10,foley11}, but which then exploded as a SN a few years
later \citep{mauerhan13}.  Spectra of pre-SN outbursts showed some
very fast material in SN~2009ip with speeds as high as 7,000-10,000 km
s$^{-1}$ \citep{smith10,foley11,pastorello13}.  An important point,
though, is that this fast material was only seen in absorption along
the line of sight, whereas the main emission line core was relatively
narrow, indicating outflow speeds for the bulk of the material of 600
km s$^{-1}$.  This situation, with the bulk of the mass moving more
slowly and a fraction of the ejecta accelerated to very high speeds,
is strikingly similar to what we see in echoes of $\eta$ Car and in
its fast Outer Ejecta \citep{smith08}.  This provides yet another
empirical link between LBVs and the progenitors of some SNe~IIn, and
provides interesting clues to the possibly similar mechanisms of
pre-SN eruptive mass loss.  The existence of this fast ejecta is a
strong constraint for any physical model of the eruption mechanism.

\section{Summary and Conclusions}

This paper presents spectra of light echoes from $\eta$~Carinae that
correspond to the time period of the main plateau of the eruption
during the late 1840s through the 1850s.  The full spectral evolution,
photometry, and other details of this echo will be discussed in a
forthcoming paper (S18).  Here we focus on one important aspect that
is significant on its own, which is the discovery of extremely broad
emission wings of H$\alpha$ that represent the fastest material ever
detected in an LBV-like eruption.  The main results from this work are
summarized as follows.

1.  In addition to a relatively narrow (600 km s$^{-1}$) line core,
H$\alpha$ displays extremely broad wings in emission, reaching to
approximately $-$10,000 km s$^{-1}$ to the blue and $+$20,000 km
s$^{-1}$ or more on the red wing.

2.  We demonstrate that the broad wings are not instrumental.  They
are not present in a nearby field star included in the same slit, and
moreover, the same broad wings are seen in spectra obtained with
different instruments on different telescopes as well as two different
gratings on the same spectrograph.  The strength of the broad wings
changes with time, and the broad emission is not seen in a different
echo that traces earlier epoch in the eruption seen from a similar
direction.  Correcting for the telluric B-band absorption, the red
wing clearly extends to $+$20,000 km s$^{-1}$ or more.

3.  The shape of the broad wings is inconsistent with electron
scattering wings, and we argue that the broad emission must trace
Doppler shifts from bulk expansion velocities.  Therefore, these are
the highest outflow speeds discovered yet in an LBV or any
non-terminal eruptive transient.

4.  The high velocities are too fast for any previously conceived
escape velocity in the system, but similar to outflow speeds from
accreting compact objects or expansion speeds of SN ejecta.  The
expanding material probably does not arise in a steady wind, but
instead likely indicates a shock-accelerated outflow.

5.  The broad wings are seen in echo spectra that view $\eta$~Car from
the equator, so these high speeds are probably not indicative of a
polar jet (even one from a compact object).  The high speeds in echoes
seen from the equator combined with fast polar speeds in the Outer
Ejecta seen today \citep{smith08} suggest a wide-angle explosion
rather than a highly collimated jet.

6.  The viewing angle of this echo could be special, however, since it
is looking from a similar direction as the ``S Condensation'' in the
Outer Ejecta.  This is also a special direction in the present day
binary system, since it is situated preferably to view the wide
companion plunge into a putative common envelope, for example (see
text).

7.  Regardless of the physical interpretation, the dual presence of
fast and slow speeds (10,000-20,000 km s$^{-1}$ and 600-1000 km
s$^{-1}$, respectively) point to CSM interaction at work in the
eruption.  They are also similar to slow and high velocities seen in
spectra of the eruptive progenitor of SN~2009ip. Therefore, the high
velocities in $\eta$ Car provide yet another interesting possible link
between LBVs and SNe~IIn.

\section*{Acknowledgements}

\scriptsize 

We thank an anonymous referee for a careful reading of the manuscript
and constructive comments.  We acknowledge contributions of additional
collaborators who helped with imaging observations to discover and
monitor light echoes, as well as for discussions and contributions to
proposals for telescope time for this and related projects: the
Carnegie Supernova Project, Alejandro Clocchiatti, Steve Margheim,
Doug Welch, and Nolan Walborn.  In particular, Nolan Walborn provided
helpful comments on the manuscript just weeks before he passed away,
which occurred while this paper was under review.  His contributions
to massive star research have been tremendous, and his unique insight
will be sorely missed.

NS's research on Eta Carinae's light echoes and related LBV-like
eruptions received support from NSF grants AST-1312221 and
AST-1515559.  Support for JLP is provided in part by FONDECYT through
the grant 1151445 and by the Ministry of Economy, Development, and
Tourism’s Millennium Science Initiative through grant IC120009,
awarded to The Millennium Institute of Astrophysics, MAS.  DJJ
gratefully acknowledges support from the National Science Foundation,
award AST-1440254.

This paper includes data gathered with the 6.5m Magellan Telescopes
located at Las Campanas Observatory, Chile.  Based, in part, on
observations obtained at the Gemini Observatory, which is operated by
the Association of Universities for Research in Astronomy, Inc., under
a cooperative agreement with the NSF on behalf of the Gemini
partnership: the National Science Foundation (United States), the
National Research Council (Canada), CONICYT (Chile), Ministerio de
Ciencia, Tecnolog\'{i}a e Innovaci\'{o}n Productiva (Argentina), and
Minist\'{e}rio da Ci\^{e}ncia, Tecnologia e Inova\c{c}\~{a}o (Brazil)
(Program GS-2014B-Q-24).

\end{document}